\documentclass[a4paper,10pt]{article}
\usepackage[utf8]{inputenc}
\usepackage{amsmath,amssymb}
\usepackage{graphicx}
\usepackage{natbib}
\usepackage{hyperref}

\newcommand{\RR}{\mathbb{R}}
\newcommand{\EE}{\mathbb{E}}
\newcommand{\Var}{\mathbb{V}\mathrm{ar}}
\newcommand{\Cov}{\mathbb{C}\mathrm{ov}}

\newcommand{\varfit}{\mathrm{var}_{\mathrm{fit}}}
\newcommand{\varres}{\mathrm{var}_{\mathrm{res}}}
\newcommand{\btheta}{\boldsymbol{\theta}}
\newcommand{\bbeta}{\boldsymbol{\beta}}
\newcommand{\bgamma}{\boldsymbol{\gamma}}
\newcommand{\bpsi}{\boldsymbol{\psi}}
\newcommand{\bphi}{\boldsymbol{\phi}}
\newcommand{\by}{\mathbf{y}}
\newcommand{\bx}{\mathbf{x}}
\newcommand{\bX}{\mathbf{X}}
\newcommand{\bb}{\mathbf{b}}

\newcommand{\pred}{\mathrm{pred}}
\newcommand{\dd}{\mathrm{d}}
\newcommand{\Tr}{\mathsf{T}}

\title{A note on Bayesian R-squared for generalized additive mixed models}
\author{Abdollah Jalilian\\
    Lancaster Ecology and Epidemiology Group, Lancaster University, UK\\
    Aki Vehtari\\
    Department of Computer Science, Aalto University, Espoo, Finland\\
    and \\
    Luigi Sedda \\
    Lancaster Ecology and Epidemiology Group, Lancaster University, UK}

\begin{document}

\maketitle

\begin{abstract}
We present a novel Bayesian framework to decompose the posterior predictive variance in a fitted Generalized Additive Mixed Model (GAMM) into explained and unexplained components. This decomposition enables a rigorous definition of Bayesian $R^{2}$. We show that the new definition aligns with the intuitive Bayesian $R^{2}$ proposed by Gelman, Goodrich, Gabry, and Vehtari (2019) [\emph{The American Statistician}, \textbf{73}(3), 307–309], 
but extends its applicability to a broader class of models. Furthermore, we introduce a partial Bayesian $R^{2}$ to quantify the contribution of individual model terms to the explained variation in the posterior predictions.\\

\noindent%

\noindent%
\end{abstract}

\section{Introduction}

The coefficient of determination, commonly denoted as $R^2$, is a widely used metric in linear models to assess how well the model explains the variability in the data. Within the framework of frequentist statistical inference, $R^2$ has been adapted to extend its applicability to generalized linear models (GLMs). Notable approaches include definitions based on likelihood ratio \citep{magee1990r}, Kullback-Leibler divergence \citep{cameron1997r}, discrimination \citep{tjur2009coefficients}, and variance function \citep{zhang2017coefficient}. However, defining $R^2$ for models with random effects, such as generalized linear mixed models (GLMMs), is not straightforward \citep{piepho2019coefficient}. As discussed in \cite{nakagawa2013general}, \cite{jaeger2017r} and \cite{rights2019quantifying}, there is no consensus in the statistical literature on the best definition of  $R^{2}$ for GLMMs and existing definitions face theoretical and practical challenges, such as incomplete partitioning of variance and inconsistent implementation across different software. Because of these issues, finding a universally accepted and comprehensive definition of $R^2$ for the broader class of generalized additive models (GAMMs), which includes both GLMs and GLMMs as special cases, remains an unsolved problem.

Within the Bayesian statistical inference paradigm, \cite{gelman2006bayesian}  developed a Bayesian $R^2$ measure by calculating the expected value of the residual sum of squares with respect to the posterior distribution of the vector of all unknowns in the model, including parameters and random effects. This approach aligns with the simple idea of using the posterior mean of the classical definition of $R^{2}$. However, as demonstrated by \cite{gelman2019r} using a toy example, this  Bayesian version of $R^2$ can produce values greater than one, leading to challenges in both application and interpretation. To resolve this issue, \cite{gelman2019r} proposed an alternative intuitive definition of Bayesian $R^2$, where sums of squares in the classical $R^2$ are replaced by their corresponding expected values based on observations generated from the predictive posterior distribution. This new definition ensures that the Bayesian $R^{2}$ remains within the interval $[0, 1]$ and offers a more interpretable measure of a model's explanation power in the Bayesian framework.

The definition of the Bayesian $R^2$ by  \cite{gelman2019r} is intuitive and does not rely on the traditional approach of decomposing total variations and calculating the ratio of explained to total variations. In this paper, we introduce a decomposition of the expected variation in posterior predictions into explained and unexplained components. This decomposition allows us to naturally define the Bayesian $R^2$ as the ratio of explained variations to total variations, resulting in a more consistent and interpretable definition for the Bayesian $R^2$. We show that this definition aligns with the one proposed by \cite{gelman2019r}.

Although the Bayesian $R^2$ by  \cite{gelman2019r} can be applied to any generalized linear model (GLM), it is currently only implemented for GLMs with Gaussian and binomial distributions in the \texttt{R} \citep{Rcore2024} package \texttt{rstantools} \citep{rstantools2024}. Our definition of Bayesian $R^2$  extends to a broader class of GAMMs and is implemented for various commonly used distributions.

The remainder of the paper is structured as follows. First, \autoref{sec:background} provides an overview of GAMMs, the classical $R^2$ and a naive Bayesian version of $R^2$. Then, \autoref{sec:bayesianR2} presents the novel decomposition of posterior predictive variations and the new definition of the Bayesian $R^2$. In~\autoref{sec:partialR2}, we introduce the concept of partial Bayesian $R^2$, which quantifies the contributions of different model terms to the posterior predictive variations. The application of the proposed $R^2$ to a simulated dataset is demonstrated in~\autoref{sec:application}. Finally,  in~\autoref{sec:conclusion} offers some concluding remarks.

\section{Background}
\label{sec:background}

Consider a GAMM for the observations,  $(y_i, \bx_i)$, $i=1,\ldots, n$, where $y_i$ represents the response (dependent or outcome) variable taking values in the space $\mathcal{Y}\subset\RR$ and
\[
  \bx_i = \left(x_{i,1}, \ldots, x_{i, m_1}, z_{i,1}, \ldots, z_{i,m_2}, u_{i,1}, \ldots, u_{m_3} \right)^{\Tr}
\]
is the vector of $m=m_1 + m_2 + m_3$ covariates (independent or explanatory variables). In this model, the conditional mean of the response variable is expressed as
\begin{equation}\label{eq:condmean}
  \mu(\bx_i, \bbeta, \bb, \bgamma) = \EE\left[ y_i | \bx_i,\bbeta, \bb, \bgamma \right] = g^{-1}\left( \eta(\bx_i,\bbeta,\bb, \bgamma) \right),
\end{equation}
where $g:\RR\to \RR$ is a known link function, and
\[
  \eta(\bx_i,\bbeta,\bb, \bgamma) = \beta_0 + \sum_{j=1}^{m_1} \beta_{j} x_{i,j} + \sum_{j=1}^{m_2} b_j z_{i, j} + \sum_{j=1}^{m_3} \mathsf{f}_j(u_{i, j}, \bgamma_j)
\]
is a linear combination of fixed effects $\bbeta=(\beta_0,\beta_1, \ldots,\beta_{m_1})$, random effects $\bb=(b_1,\ldots,b_{m_2})$ and smooth functions $\mathsf{f}_1, \ldots, \mathsf{f}_{m_3}$. Here, each smooth function $\mathsf{f}_j(u, \bgamma_j)$ is assumed to be given by
\[
  \mathsf{f}_j(u, \bgamma_j) = \sum_{l=1}^{k_j} \gamma_{j,l} h_{j,l}(u),
\]
where $h_{j,l}(u)$ are known spline basis functions chosen to have convenient properties, and $\gamma_{j,l}$ are unknown coefficients that need to be estimated. Often, the basis dimension $k_j$ is fixed at a sufficiently large size and the smoothness of $\mathsf{f}_j$ function is controlled by imposing a constraint on a \emph{wiggliness} measure such as
\begin{equation}\label{eq:wiggliness}
  \Omega_j(\bgamma_j) = \bgamma_{j}^{\Tr} S_{j} \bgamma_{j},
\end{equation}
where $S_j$ is a semi-definite matrix of known constants \citep{wood2017generalised}.

Moreover, the conditional distribution of the random effects $\bb$ and the response vector $\by=(y_1,\ldots,y_n)^{\Tr}$ are given by
\begin{equation}\label{eq:conddist}
 \left\{ \begin{array}{l}
          \bb | \bX,\bpsi \, \sim\, q(\bb|\bX,\bpsi) \vspace*{0.2cm} \\
          \by | \bX, \bbeta, \bb, \bgamma, \bphi \, \sim \, \prod_{i=1}^{n} f(y_i|\bx_i, \bbeta, \bb, \bgamma, \bphi)
         \end{array}\right.,
\end{equation}
where $\bX=[\bx_{1}^{\Tr},\ldots, \bx_{n}^{\Tr}]^{\Tr}$ is the matrix of covariates, $q(\bb|\bX,\bpsi)$ is the probability density function of random effects with parameter vector $\bpsi$, and  $f(y|\bx, \bbeta, \bb, \bgamma, \bphi)$ denote the probability density (mass) function of a distribution belonging to the exponential family of distributions which depends on the coefficients $\bbeta$, $\bb$ and $\bgamma$ and possibly nuisance parameters $\bphi$. The conditional variance
\begin{equation}\label{eq:condvar}
  \sigma^{2}(\bx_i,\bbeta, \bb, \bgamma, \bphi) = \Var[y_i|\bx_i,\bbeta, \bb, \bgamma, \bphi] = V\left( \mu(\bx_i, \bbeta, \bb, \bgamma), \bphi\right)
\end{equation}
depends upon the conditional mean~\eqref{eq:condmean} through the variance function $V(\,\cdot\,, \bphi):\RR\to\RR_{+}$ as well as upon extra parameters $\bphi$ \citep[see][Chapter~7]{mcculloch2008generalized}. \autoref{tab:glmmexamp} provides examples of common distributions used for the response variable, along with their corresponding link and variance functions.

\begin{table}
\caption{Examples of common conditional distributions for the response variable in a GAMM and their corresponding response space $\mathcal{Y}$, link function $g(x)$, variance function $V(x, \bphi)$ and dispersion parameter $\bphi$.}
\label{tab:glmmexamp}
\centering
 \begin{tabular}{l|c|c|c|c}
 distribution & $\mathcal{Y}$ & $g(x)$ & $V(x, \bphi)$ & dispersion parameter $\bphi$\\\hline
 Gaussian  & $\RR$ & $x$ & $\frac{1}{\phi}$ & $\phi = 1/\text{variance}$\\
 Gamma     & $\RR_+$ & $1/x$ & $\frac{x^2}{\phi}$ & $\phi=\text{shape}$ \\
 Inverse Gaussian & $\RR_+$ & $1/x^2$ & $\frac{x^3}{ \phi}$ & $\phi=1/\text{scale}$\\
 Beta & $[0,1]$ & $\log \frac{x}{1 - x}$ & $\frac{x(1 - x)}{(1+\phi)}$ & $\phi=\text{shape}_1+\text{shape}_2$ \\
 Bernoulli & $\{0,1\}$ & $\log \frac{x}{1 - x}$ & $x(1 - x)$ & --- \\
 Poisson & $\mathbb{Z}_+$ &  $\log(x)$ &  $x$ & --- \\
 Negative binomial & $\mathbb{Z}_+$ & $\log(x)$ & $x \left(1 + \frac{x}{\phi} \right)$ & $\phi=\text{shape}$\\
\end{tabular}
\end{table}

\subsection{Parameter estimation}

Let $\btheta=(\bbeta, \bgamma, \bpsi, \bphi)$ denote the vector of all model parameters. The likelihood function for the response $\by$ and covariates $\bX$ is given by
\[
  p(\by|\bX,\btheta) = \int \prod_{i=1}^{n} f(y_i|\bx_i, \bbeta, \bb, \bgamma, \bphi) q(\bb|\bX,\bpsi) \dd \bb.
\]
In frequentist statistical inference, the maximum likelihood estimate of $\btheta$ is obtained by maximizing a penalised version of the above likelihood function with respect to $\btheta$; i.e.,
\[
  \widehat{\btheta} = \arg\max_{\btheta} \left( p(\by|\bX,\btheta) + \sum_{j=1}^{m_3} \lambda_j \Omega_{j}(\bgamma_j) \right),
\]
where $\Omega_{j}(\bgamma_j)$ is a wiggliness measure as in \eqref{eq:wiggliness} and $\lambda_1,\ldots,\lambda_{m_3}$ are smoothing parameters \citep{wood2017generalised}.

In Bayesian inference \citep{gelman2013bayesian}, model parameters $\btheta$ are regarded as random variables and their uncertainty is quantified using a prior probability distribution with density function $\pi(\btheta)=\pi(\bbeta)\pi(\bgamma)\pi(\bpsi)\pi(\bphi)$. Random effects $\bb$ are also considered as unknown parameters of the model, with their prior distribution given by $q(\bb|\bX, \bpsi)$. Moreover, improper priors such as
\[
  \pi(\bgamma) \propto \exp\left( -\sum_{j=1}^{m_3} \lambda_j \Omega_{j}(\bgamma_j) / \alpha_j \right)
\]
can be defined to control the wiggliness of $\mathsf{f}_j$'s, where $\lambda_1,\ldots,\lambda_{m_3}$ are smoothing parameters and $\alpha_1,\ldots,\alpha_{m_3}$ are hyper parameters \citep{wood2013straightforward}. The posterior density
\begin{equation}\label{eq:posteiordens}
  \pi(\bb, \btheta|\by,\bX) = \frac{\prod_{i=1}^{n} f(y_i|\bx_i, \bbeta, \bb, \bgamma, \bphi) q(\bb|\bX,\bpsi) \pi(\btheta)}{p(\by|\bX)}.
\end{equation}
is used to make any inference about $\bb$ and $\btheta$. For example, the posterior mean can be used to estimate $\bb$ and $\btheta$.

Efficient computational implementations of the above-mentioned frequentist and Bayesian estimation approaches of the model parameters in \texttt{R} are available in packages such as \texttt{gamm4} \citep{gamm42020} and \texttt{rstanarm} \citep{rstanarm2024}, respectively.

\subsection{Classical coefficient of determination}

In the case of a linear model (identity link function) with no random effects or smooth terms, and a conditional Gaussian distribution for the response variable, the conditional mean~\eqref{eq:condmean} and variance~\eqref{eq:condvar} are reduced to
\begin{align*}
  \mu(\bx_i, \bbeta) &= \EE[y_i|\bx_i,\bbeta] = \eta(\bx_i,\bbeta) = \bx_{i}^{\Tr} \beta, \\
  \sigma^{2}(\bx_i,\bbeta, \sigma^2) &= \Var[y_i|\bx_i,\bbeta, \sigma^2] \equiv \sigma^{2}.
\end{align*}
In a frequentist framework, the response $y_i$ can be expressed as
\[
   y_i = \overbrace{\mu(\bx_i, \widehat{\bbeta})}^{\mathrm{model}} \, + \, \overbrace{e_i(\bx_i, \widehat{\bbeta})}^{\mathrm{residual}}, \quad i=1,\ldots,n,
\]
where $e_i(\bx_i, \widehat{\bbeta}) = y_i - \mu(\bx_i, \widehat{\bbeta})$ and  $\widehat{\bbeta}=(\bX^{\Tr} \bX)^{-1} \bX^{\Tr} \by$ is the least squares and maximum likelihood estimate of $\bbeta$. Moreover, the total sum of squares of deviations for the response vector $\by$,
\[
  \mathrm{TSS} = \sum_{i=1}^{n} (y_i - \bar{y})^2
\]
is decomposed to
\begin{equation}\label{eq:TSSdecomp}
    \mathrm{TSS} = \mathrm{RSS}(\widehat{\bbeta}) + \mathrm{ESS}(\widehat{\bbeta}),
\end{equation}
where
\begin{equation}\label{eq:ssreslm}
  \mathrm{RSS}(\widehat{\bbeta}) = \sum_{i=1}^{n} \left( y_i - \mu(\bx_i, \widehat{\bbeta}) \right)^2 
  = \sum_{i=1}^{n} \left( e_i(\bx_i, \widehat{\bbeta})  - \frac{1}{n}\sum_{j=1}^{n} e_j(\bx_j, \widehat{\bbeta}) \right)^2
\end{equation}
is the residual sums of squares, and
\begin{equation}\label{eq:ssreg}
  \mathrm{ESS}(\widehat{\bbeta}) = \sum_{i=1}^{n} \left( \mu(\bx_i, \widehat{\bbeta}) - \bar{y} \right)^2 = \sum_{i=1}^{n} \left( \mu(\bx_i, \widehat{\bbeta}) - \frac{1}{n} \sum_{j=1}^{n} \mu(\bx_j, \widehat{\bbeta}) \right)^2
\end{equation}
is the explained sum of squares. The last two equations in~\eqref{eq:ssreslm} and~\eqref{eq:ssreg} hold because $\sum_{i=1}^{n} \mu(\bx_i, \widehat{\bbeta}) / n= \bar{y}$ and hence $\sum_{j=1}^{n} e_j(\bx_j, \widehat{\bbeta}) = 0$.

This decomposition breaks down the total variations in the observed response vector $\by$ into two parts: the amount of variation explained by the covariates' matrix $\bX$ through the specified model and the residual unexplained variation. Thus, the coefficient of determination
\begin{equation}\label{eq:classicalr2}
  R_{\mathrm{classic}}^{2} = \frac{\mathrm{ESS}(\widehat{\bbeta})}{\mathrm{TSS}} = \frac{\mathrm{ESS}(\widehat{\bbeta})}{\mathrm{RSS}(\widehat{\bbeta}) + \mathrm{ESS}(\widehat{\bbeta})}.
\end{equation}
quantifies the proportion of variations in the response vector $\by$ that is explained by the considered linear model.

In the case of a GAMM model defined by distribution structure~\eqref{eq:conddist} and conditional mean~\eqref{eq:condmean},  unlike the parameters $\btheta$, random effects $\bb$ are unobservable random quantities. As a result, the conditional mean $\mu(\bx_i, \bbeta, \bb, \bgamma)$ and the residual
\[
  e_i(\bx_i, \bbeta, \bb, \bgamma)= y_i - \mu(\bx_i, \bbeta, \bb, \bgamma)
\]
are not directly computable. This makes it challenging to define computable equivalents for the residual sum of squares~\eqref{eq:ssreslm} and the explained sum of squares~\eqref{eq:ssreg} \citep{jaeger2017r, piepho2019coefficient}.

\subsection{Bayesian alternative}

In the Bayesian approach, the random effects $\bb$ are treated as parameters, allowing the residual sum of squares~\eqref{eq:ssreslm} and the explained sum of squares~\eqref{eq:ssreg} to be expressed in terms of the conditional mean $\mu(\bx_i, \bbeta, \bb, \bgamma)$ and the residual $e_i(\bx_i, \bbeta, \bb, \bgamma)$ instead of $\mu(\bx_i, \widehat{\bbeta})$ and $e_i(\bx_i, \widehat{\bbeta})$. In this approach, $\mathrm{RSS}(\bbeta, \bb, \bgamma)$ and $\mathrm{ESS}(\bbeta, \bb, \bgamma)$ are considered random quantities, while the total sum of squares $\mathrm{TSS}$ remains a constant. Within this framework, the coefficient of determination \eqref{eq:classicalr2} can be replaced by the Bayesian version~\citep{gelman2006bayesian}
\[
  R_{\mathrm{Bayes}}^{2} = \EE\left[ R_{\mathrm{classic}}^{2}  \big| \by, \bX \right]
  = \frac{\EE\left[ \mathrm{ESS}(\bbeta, \bb, \bgamma) \big| \by, \bX \right]}{\mathrm{TSS}},
\]
where the numerator is the posterior mean of
\[
  \mathrm{ESS}(\bbeta, \bb, \bgamma) =  \sum_{i=1}^{n} \left( \mu(\bx_i, {\bbeta}, \bb, \bgamma) - \frac{1}{n} \sum_{j=1}^{n} \mu(\bx_j, {\bbeta}, \bb, \bgamma) \right)^2;
\]
i.e., expectation of $\mathrm{ESS}(\bbeta, \bb, \bgamma)$ with respect to the posterior density~\eqref{eq:posteiordens}.

However, as demonstrated in a toy example by \cite{gelman2019r}, the posterior mean $\EE\left[ \mathrm{ESS}(\bbeta, \bb, \bgamma) \big| \by, \bX \right]$ can be larger than the total sum of squares $\mathrm{TSS}$. This means that $R_{\mathrm{Bayes}}^2$ can exceed one, making it impossible to interpret as the proportion of variations in the data explained by the model. In~\autoref{sec:bayesianR2} we use the posterior predictive density to provide a similar decomposition of the total sum of squares and a definition of $R^{2}_{\mathrm{Bayes}}$ without the above limitation.

\section{Bayesian coefficient of determination}
\label{sec:bayesianR2}

The  posterior predictive density
\begin{align}
  p^{\pred}(\tilde{\by}|\by, \bX) &= \int \left( \prod_{i=1}^{n} f(\tilde{y}_i|\bx_i, \bbeta, \bphi, \tilde{\bb}) \right) \pi(\tilde{\bb}, \btheta|\by,\bX) \dd \tilde{\bb} \dd \btheta \label{eq:predpostdens}
\end{align}
represents the uncertainty about future observation $\tilde{\by}$ for the response vector given the observed data $\by$ and $\bX$, after accounting for the uncertainty in random effects $\tilde{\bb}$ and model parameters $\btheta$ \citep[see][Section~6.3]{gelman2013bayesian}.  Note that the predictive posteior~\eqref{eq:predpostdens} assumes that given $\bX$ and $\btheta$, the potential future response vector $\tilde{\by}$ is conditionally independent of the observed response vector $\by$; i.e.,
\begin{equation*}\label{eq:condindep}
p(\tilde{\by}, \by| \bX, \btheta) = p(\tilde{\by}|\bX,\btheta) p(\by|\bX,\btheta).
\end{equation*}

\subsection{Decomposition of posterior predictive variations}

If $\tilde{\by}$ is generated from the posterior predictive distribution $p^{\pred}(\tilde{\by}|\by, \bX)$, it represents a potential new response vector with its associated random effects and parameter vector $(\tilde{\bb}, \btheta) \, \sim\, \pi(\tilde{\bb}, \btheta|\by,\bX)$. In other words, $\tilde{\by}$ imitates what new data might look like based on the considered model and the observed data $\by$ and $\bX$. Thus, variation of $\tilde{\by}$ reflects the plausible variability in future observations, stemming not only from the randomness in the observed data but also from the uncertainty in the model's structure.

We consider the total sum of squares of posterior predictive deviations defined as
\begin{align*}
  \widetilde{\mathrm{TSS}}(\tilde{\bb}, \btheta|\by,\bX) &= \EE\left[ \sum_{i=1}^{n} \left( \tilde{y}_i - \bar{\tilde{y}} \right)^2 \big| \by, \bX, \tilde{\bb}, \btheta \right],
\end{align*}
where $\bar{\tilde{y}} = \sum_{i=1}^{n} \tilde{y}_i / n$ is the mean of the future response values. We define the posterior predictive residual
\[
  \tilde{e}_i(\bx_i, \bbeta, \tilde{\bb}, \bgamma) = \tilde{y}_i - \mu(\bx_i, \bbeta, \tilde{\bb}, \bgamma)
\]
as differences between future response value $\tilde{y}_i$ and its corresponding conditional predictive mean $\mu(\bx_i,\bbeta, \tilde{\bb},\bgamma)$. Similar to~\eqref{eq:ssreslm}, we define the posterior predictive residual sum of squares as
\[
  \widetilde{\mathrm{RSS}}(\tilde{\bb}, \btheta|\by,\bX) =  \EE\left[ \sum_{i=1}^{n} \left( \tilde{e}_i(\bx_i, \bbeta, \tilde{\bb},\bgamma) - \frac{1}{n} \sum_{j=1}^{n} \tilde{e}_j(\bx_j, \bbeta, \tilde{\bb},\bgamma) \right)^2 \big| \by, \bX, \tilde{\bb}, \btheta \right].
\]
The quantity $\widetilde{\mathrm{TSS}}(\tilde{\bb}, \btheta|\by,\bX)$ captures the overall deviation of the future response values from their mean, providing insight into the variation in potential future responses based on the considered model and available data $\by$ and $\bX$. On the other hand, $\widetilde{\mathrm{RSS}}(\tilde{\bb}, \btheta|\by,\bX)$ measures the sum of squares of deviations for differences between future response values and their corresponding conditional means, representing the residual unexplained variations.

Analogous to the decomposition~\eqref{eq:TSSdecomp}, $\widetilde{\mathrm{TSS}}(\tilde{\bb}, \btheta|\by,\bX)$ can be decomposed as (see~\autoref{sec:sstdecomp} for details)
\begin{equation}\label{eq:bayessstdecomp}
   \widetilde{\mathrm{TSS}}(\tilde{\bb}, \btheta|\by,\bX) =  \widetilde{\mathrm{RSS}}(\tilde{\bb}, \btheta|\by,\bX) +  \widetilde{\mathrm{ESS}}(\tilde{\bb}, \btheta),
\end{equation}
where, similar to~\eqref{eq:ssreg}, the explained sum of squares
\begin{align*}
  \widetilde{\mathrm{ESS}}(\tilde{\bb}, \btheta|\bX) &= \sum_{i=1}^{n} \left( \mu(\bx_i,\bbeta, \tilde{\bb}, \bgamma) - \frac{1}{n} \sum_{j=1}^{n} \mu(\bx_j,\bbeta, \tilde{\bb}, \bgamma) \right)^2
\end{align*}
captures variations in the conditional mean~\eqref{eq:condmean}. Unlike $\widetilde{\mathrm{TSS}}(\tilde{\bb}, \btheta|\by,\bX)$ and $\widetilde{\mathrm{RSS}}(\tilde{\bb}, \btheta|\by,\bX)$, $\widetilde{\mathrm{TSS}}(\tilde{\bb}, \btheta|\bX)$ does not depend on the observed response vector $\by$ and only represents variations explained by the considered model over different rows of the covariate matrix $\bX$. This decomposition means on average, the total variation in the potential future response vectors can be partitioned into two sources: variation due to the specified conditional mean~\eqref{eq:condmean} and residual unexplained variation.

\subsection{New definition of Bayesian $R^2$}

Building on the same idea and structure as the classical $R^2$ in \eqref{eq:classicalr2}, we define the Bayesian $R^{2}$ for GAMMs as
the quantity
\[
  R^{2}_{\mathrm{Bayes}}(\tilde{\bb}, \btheta|\by,\bX) = \frac{ \widetilde{\mathrm{ESS}}(\tilde{\bb}, \btheta | \bX) }{\widetilde{\mathrm{TSS}}(\tilde{\bb}, \btheta|\by,\bX)}.
\]
This definition aligns with the Bayesian $R^{2}$ proposed by \cite{gelman2019r},
\[
  R^{2}_{\mathrm{Bayes}} = \frac{\varfit}{\varres + \varfit},
\]
where $\varfit =  \widetilde{\mathrm{ESS}}(\tilde{\bb}, \btheta|\bX) / (n-1)$,
\[
 \varres = \frac{1}{n} \sum_{i=1}^{n} \sigma^{2}(\bx_i, \bbeta, \bb, \bgamma, \bphi).
\]
and $\sigma^{2}(\bx_i, \bbeta, \bb, \bgamma, \bphi)$ is the conditional variance~\eqref{eq:condvar} (see~\autoref{sec:bayesRsq} for details).

To estimate $R^{2}_{\mathrm{Bayes}}(\tilde{\bb}, \btheta|\by,\bX)$, we can generate  samples $(\tilde{\bb}_{1}, \btheta_1), \ldots, (\tilde{\bb}_{L}, \btheta_{L}) \, \sim\, \pi(\tilde{\bb}, \btheta|\bX, \by)$ from the posterior distribution~\eqref{eq:posteiordens} to obtain corresponding samples $R^{2}_{\mathrm{Bayes}}(\tilde{\bb}_l, \btheta_l|\by, \bX)$, $l = 1, \ldots, L$. These samples can then be used to approximate the distribution or specific properties (such as the mean) of $R^{2}_{\mathrm{Bayes}}(\tilde{\bb}, \btheta|\by,\bX)$,  which is a standard approach in Bayesian statistical inference.

\section{Partial R-squared}
\label{sec:partialR2}

As proposed by \cite{nakagawa2013general} and \cite{jaeger2017r}, comparing a reduced model that includes either only fixed or only random effects with the full model that consists of both can serve as a basis for defining partial or marginal metrics of goodness of fitting.
However, the same idea can also be extended to compare any reduced model that is a simpler version of the full model by leaving out specific fixed effects, random effects or smooth terms. For example, by excluding the random effects and smoothing terms from the conditional mean~\eqref{eq:condmean}, we obtain the reduced model
\[
  \mu_0(\bx_i, \bbeta) = g^{-1}\left( \eta_0(\bx_i, \bbeta)\right)
\]
with $\eta_0(\bx_i, \bbeta) = \bx_{i,1}^{\Tr} \bbeta$.

Similar to the concept defined for linear models, we define the partial (also known as marginal or Type III) sum of squares for the terms in the full model that have been excluded in the reduced model.
This is defined as the conditional expected difference between the residual sum of squares of the full model and that of the reduced model; i.e., (see \autoref{sec:partialdecomp})
\begin{align}
  \widetilde{\mathrm{ESS}}_{1} & (\tilde{\bb}, \btheta | \bX) =  \widetilde{\mathrm{RSS}}_0(\tilde{\bb}, \btheta | \by, \bX) - \widetilde{\mathrm{RSS}}(\tilde{\bb}, \btheta | \by, \bX) \label{eq:ssresreduced} \\
  &= \nonumber \sum_{i=1}^{n} \left( \left( \mu(\bx_i, \bbeta, \tilde{\bb}, \bgamma) - \mu_0(\bx_i, \bbeta) \right) - \frac{1}{n} \sum_{j=1}^{n} \left( \mu(\bx_j, \bbeta, \tilde{\bb}, \bgamma) - \mu_0(\bx_j, \bbeta) \right) \right)^2.
\end{align}
Here, $\widetilde{\mathrm{ESS}}_{1}(\tilde{\bb}, \btheta | \bX)$ represents the variation attributable solely to the difference between the full and reduced models.
In other words, $\widetilde{\mathrm{ESS}}_{1}(\tilde{\bb}, \btheta | \bX)$ can be interpreted as the sum of squares of deviation for including excluded terms (i.e., random effect) in the reduced model.

Equation \eqref{eq:ssresreduced} implies that the posterior predictive residual sum of squares for the reduced model, $\widetilde{\mathrm{RSS}}_0(\tilde{\bb}, \btheta | \by, \bX)$, can be decomposed into two non-negative components:
\begin{itemize}
 \item one component, $\widetilde{\mathrm{RSS}}(\tilde{\bb}, \btheta | \by, \bX)$, is the posterior predictive residual sum of squares for the full model, and
 \item the other component, $\widetilde{\mathrm{ESS}}_{1}(\tilde{\bb}, \btheta |  \bX)$, captures the sum of squared deviations caused by the inclusion of excluded terms in the reduced model.
\end{itemize}
Thus, the ratio
\[
  \frac{\widetilde{\mathrm{ESS}}_{1}(\tilde{\bb}, \btheta |  \bX)}{ \widetilde{\mathrm{RSS}}(\tilde{\bb}, \btheta | \by, \bX) + \widetilde{\mathrm{ESS}}_{1}(\tilde{\bb}, \btheta |  \bX) }
\]
represents the amount of total posterior predictive variation not explained by the reduced model that can be attributed to the excluded terms from the full model. This ratio can be considered as a partial Bayesian $R^{2}$ to quantify the contribution of the excluded terms.

Note that the above ratio is different than the ratio
\[
  \frac{\widetilde{\mathrm{ESS}}_{1}(\tilde{\bb}, \btheta | \bX)}{\widetilde{\mathrm{TSS}}(\tilde{\bb}, \btheta | \by, \bX)}
\]
which measures the proportion of the total posterior predictive variation explained by the excluded terms in the full model.

The same approach can be repeated for any other term of the model by removing it from the conditional mean~\eqref{eq:condmean} to obtain the corresponding partial Bayesian $R^{2}$.

\section{Application}
\label{sec:application}

To illustrate the applicability of the introduced $R^{2}_{\mathrm{Bayes}}$ and its partial variant, we simulated a dataset with $n=200$ observations from a negative binomial model with the shape (size) parameter $\phi=2$. The conditional mean is defined as
\[
  \mu(\bx_i, \bbeta, \bb, \bgamma) = \exp\left(\beta_0 + \beta_1 x_{i,1} + b_{1}z_{i, 1} + b_{2} z_{i,2} + \mathsf{f}_1(u_{i,1}) \right),
\]
where the covariates $x_{i,1}$ and $u_{i,1}$, for $i=1, \ldots, n$, are drawn from a uniform distribution on the interval (0, 1), with $\beta_0=3$ and $\beta_1=2$.
The smooth function $\mathsf{f}_1(u)$ is constructed as
\[
  \mathsf{f}_1(u) =  2250\left( 20 u^{11} (1 - u)^{6} + u{^3} (1 - u)^{10} \right) - \frac{3}{4}.
\]
The random effects $b_1$ and $b_2$ are sampled from a standard normal distribution. The variables $z_{i,1}$ and $z_{i,2}$ are defined as $z_{i,1} = v_i$ and $z_{i,2} = 1 - v_i$, where $v_i$, for $i = 1, \ldots, n$, is drawn from a Bernoulli distribution with a success probability of $0.5$. This setup partitions the $n$ observations into two groups, one associated with random effect $b_1$ and the other with $b_2$.

\begin{verbatim}
n <- 200
u1 <- runif(n, 0, 1)
f1 <- 2250 * (20 * u1^11 * (1 - u1)^6 + u1^3 * (1 - u1)^10) - 3/4
b1 <- rnorm(2, 0, 1)
dat <- data.frame(x1=runif(n, 0, 1),
    z1=sample(1:2, size=n, replace=TRUE), u1=u1, f1=f1)
eta <- 3 + 2 * dat$x1 +  b1[dat$z1] + dat$f1
dat$y <- rnbinom(n, mu=exp(eta), size=2)
\end{verbatim}

We fit three models to the simulated data using the model-fitting functions from the \texttt{rstanarm} package in \texttt{R}: a model with only the fixed effect (\texttt{fit0}), a model with both fixed and random effects (\texttt{fit1}), and a model including the fixed and random effects and the smooth term (\texttt{fit2}).

\begin{verbatim}
library("rstanarm")
fit0 <- stan_glm(y ~ x1, data=dat, family=neg_binomial_2)
fit1 <- stan_glmer(y ~ x1 + (1|z1), data=dat, family=neg_binomial_2)
fit2 <- stan_gamm4(y ~ x1 + s(u1), random= ~ (1|z1),
                   data = dat, family=neg_binomial_2)
\end{verbatim}

\begin{figure}
    \centering
    \includegraphics[width=0.495\textwidth]{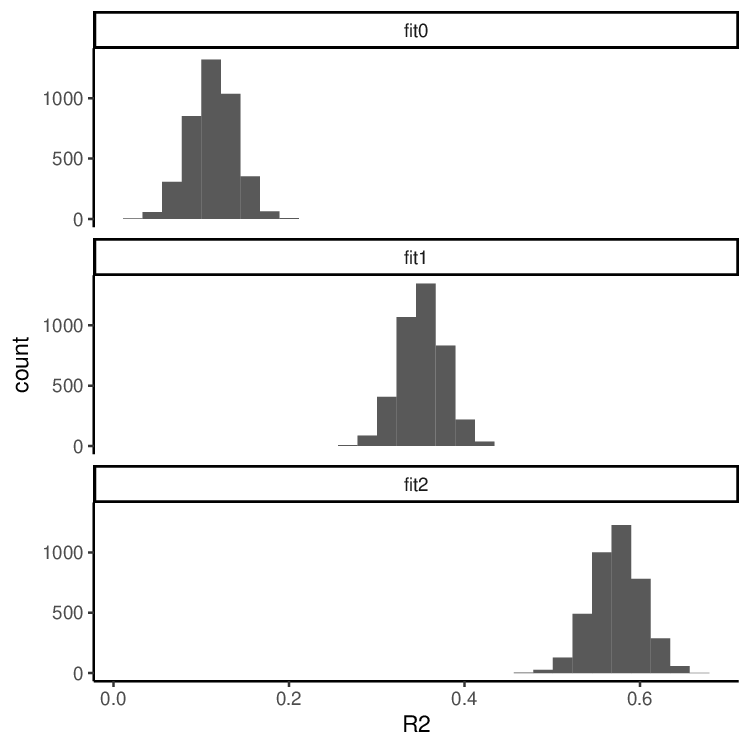}
    \includegraphics[width=0.495\textwidth]{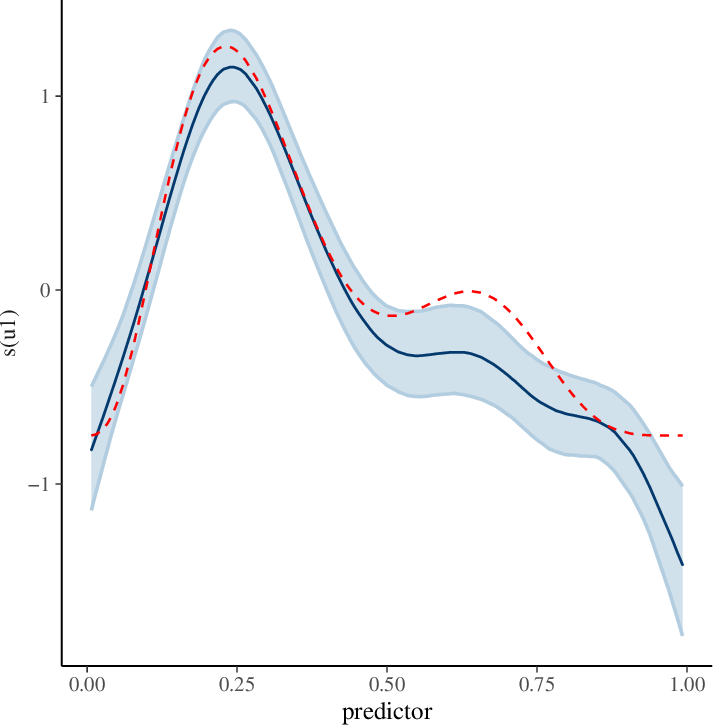}
    \caption{Left: Histograms of $R^{2}_{\mathrm{Bayes}}$ for the three models fitted to the simulated data: the model with only the fixed effect (\texttt{fit0}), the model with both fixed and random effects (\texttt{fit1}), and the model with fixed effects, random effects, and a smooth term (\texttt{fit2}). Right: The true smooth term (dashed line) compared to the estimated smooth term in the fitted model \texttt{fit2}.}
    \label{fig:fitresults}
\end{figure}

Histograms of $R^{2}_{\mathrm{Bayes}}$ for the fitted models, along with a comparison between the true smooth term (dashed line) and the estimated smooth term $\mathsf{f}_1(u)$ from the third model, are shown in~\autoref{fig:fitresults}. Despite some minor discrepancies, the estimated smooth term closely matches the true $\mathsf{f}_1(u)$.

\autoref{fig:fitresults} shows that the three fitted models explain approximately 10\%, 35\%, and 60\% of the variation in the posterior predictions, respectively. The partial Bayesian $R^{2}$ can be used to assess how much of the unexplained variation in the model \texttt{fit0} is due to the exclusion of random effects and the smooth term.
\begin{verbatim}
mean(bayes_R2_partial(fit2, term="x1"))
[1] 0.5641814
\end{verbatim}
The partial Bayesian $R^{2}$ indicates that around 55\% of the unexplained variation in the \texttt{fit0} model is attributable to the exclusion of random effects and the smooth term.

\section{Conclusion}
\label{sec:conclusion}

Despite its limitations and criticisms, such as being sensitive to sample size, not accounting for model overfitting, and potential for misleading interpretations, $R^2$ remains a popular and widely used diagnostic tool in numerous applications due to its straightforward interpretation. The concept of $R^2$ has evolved to include a broader range of statistical models beyond traditional linear models under the frequentist statistical framework. Notably, \cite{gelman2019r} proposed an intuitive definition for the Bayesian $R^2$  based on the predictive posterior distribution.

In this work, we provided a clear decomposition of the total sum of squares of posterior predictive variations into explained and unexplained components, allowing us to naturally define Bayesian $R^2$   as the ratio of explained variations to total variations. This approach provides a more consistent, applicable and interpretable measure of model fit for complex modelling
scenarios such as GAMMs. By integrating these advancements, the Bayesian $R^2$
provides a promising method for accurately assessing goodness of fit across a wide range of statistical models.

\bigskip
\begin{center}
{\large\bf SUPPLEMENTARY MATERIAL}
\end{center}

\begin{description}

\item[Implementation of Bayesian R-squared for GAMMs:] A supplementary file provides code for an updated version of the \texttt{bayes\_R2} function in the \texttt{rstantools} package, extending its applicability to GLMMs and GAMMs via the model fitting functions from the \texttt{rstanarm} package, and introduces a new function, \texttt{bayes\_R2\_partial}, for calculating partial Bayesian  $R^2$, with applications. (PDF)

\end{description}

\bibliographystyle{apalike}
\bibliography{library}

\appendix

\section{Decomposition of total sum of squares of predictive deviations}
\label{sec:sstdecomp}

Since
\begin{align*}
  \tilde{y}_i - \bar{\tilde{y}} &= \left( \left( \tilde{y}_i - \mu(\bx_i, \bbeta, \tilde{\bb}, \bgamma) \right) - \frac{1}{n} \sum_{j=1}^{n} \left( \tilde{y}_j - \mu(\bx_j, \bbeta, \tilde{\bb}, \bgamma) \right) \right) \\
  &+  \left( \mu(\bx_i, \bbeta, \tilde{\bb}, \bgamma) - \frac{1}{n} \sum_{j=1}^{n} \mu(\bx_j, \bbeta, \tilde{\bb}, \bgamma) \right)\\
  &= \left( \tilde{e}_i(\bx_i, \bbeta, \tilde{\bb}) - \frac{1}{n} \sum_{j=1}^{n} \tilde{e}_j(\bx_j, \bbeta, \tilde{\bb}) \right) \\
  &+  \left( \mu(\bx_i, \bbeta, \tilde{\bb}, \bgamma) - \frac{1}{n} \sum_{j=1}^{n} \mu(\bx_j, \bbeta, \tilde{\bb}, \bgamma) \right),
\end{align*}
we have
\begin{align*}
  \sum_{i=1}^{n} \left( \tilde{y}_i - \bar{\tilde{y}} \right)^2 &= \sum_{i=1}^{n} \left( \tilde{e}_i(\bx_i, \bbeta, \tilde{\bb}, \bgamma) - \frac{1}{n} \sum_{j=1}^{n} \tilde{e}_j(\bx_j, \bbeta, \tilde{\bb}, \bgamma) \right)^2 \\
 &+ \sum_{i=1}^{n} \left( \mu(\bx_i, \bbeta, \tilde{\bb}, \bgamma) - \frac{1}{n} \sum_{j=1}^{n} \mu(\bx_j, \bbeta, \tilde{\bb}, \bgamma) \right)^2 \\
 &+ 2  \sum_{i=1}^{n} \left( \tilde{e}_i(\bx_i, \bbeta, \tilde{\bb}) - \frac{1}{n} \sum_{j=1}^{n} \tilde{e}_j(\bx_j, \bbeta, \tilde{\bb}) \right) \left( \mu(\bx_i, \bbeta, \tilde{\bb}, \bgamma) - \frac{1}{n} \sum_{j=1}^{n} \mu(\bx_j, \bbeta, \tilde{\bb}, \bgamma) \right).
\end{align*}
The fact that
\[
  \EE\left[  \tilde{y}_i \big| \by, \bX, \tilde{\bb}, \btheta \right] = \EE\left[  \tilde{y}_i \big| \bx_i,\bbeta, \tilde{\bb}, \bgamma \right] = \mu(\bx_i,\bbeta, \tilde{\bb}, \bgamma)
\]
and hence
\[
  \EE\left[  \tilde{e}_i(\bx_i,\bbeta, \tilde{\bb}, \bgamma) \big| \by, \bX,\tilde{\bb}, \btheta \right] = 0
\]
implies that
\begin{align*}
  \EE & \left[ \left( \tilde{e}_i(\bx_i, \bbeta, \tilde{\bb}, \bgamma) - \frac{1}{n} \sum_{j=1}^{n} \tilde{e}_j(\bx_j, \bbeta, \tilde{\bb}, \bgamma) \right) \left( \mu(\bx_i, \bbeta, \tilde{\bb}, \bgamma) - \frac{1}{n} \sum_{j=1}^{n} \mu(\bx_j, \bbeta, \tilde{\bb}, \bgamma) \right) \big| \by, \bX, , \tilde{\bb}, \btheta \right] \\
  &= \left( \mu(\bx_i, \bbeta, \tilde{\bb}, \bgamma) - \frac{1}{n} \sum_{j=1}^{n} \mu(\bx_j, \bbeta, \tilde{\bb}, \bgamma) \right) \EE\left[ \left( \tilde{e}_i(\bx_i, \bbeta, \tilde{\bb}, \bgamma) - \frac{1}{n} \sum_{j=1}^{n} \tilde{e}_j(\bx_j, \bbeta, \tilde{\bb}, \bgamma) \right) \big| \bX, \tilde{\bb}, \btheta \right]  \\
  &= 0.
\end{align*}
Therefore,
\begin{align*}
   \widetilde{\mathrm{TSS}}(\tilde{\bb}, \btheta | \by, \bX) &=
  \EE\left[ \sum_{i=1}^{n} \left( \tilde{y}_i - \bar{\tilde{y}} \right)^2 | \by, \bX, \tilde{\bb}, \btheta \right] \\
  &= \EE\left[ \sum_{i=1}^{n} \left( \tilde{e}_i(\bx_i, \bbeta, \tilde{\bb}, \bgamma) - \frac{1}{n} \sum_{j=1}^{n} \tilde{e}_j(\bx_j, \bbeta, \tilde{\bb}, \bgamma) \right)^2 | \by, \bX, \tilde{\bb}, \btheta \right] \\
  &+   \sum_{i=1}^{n} \left( \mu(\bx_i, \bbeta, \tilde{\bb}, \bgamma) - \frac{1}{n} \sum_{j=1}^{n} \mu(\bx_j, \bbeta, \tilde{\bb}, \bgamma) \right)^2  \\
  &=  \widetilde{\mathrm{RSS}}(\tilde{\bb}, \btheta| \by, \bX)  + \widetilde{\mathrm{ESS}}(\tilde{\bb}, \btheta | \bX).
\end{align*}

\section{Relation with the \cite{gelman2019r} $R^2$}
\label{sec:bayesRsq}

Since
\begin{align*}
  \EE & \left[  \left( \tilde{e}_i(\bx_i, \bbeta, \tilde{\bb}, \bgamma) - \frac{1}{n} \sum_{j=1}^{n} \tilde{e}_j(\bx_j, \bbeta, \tilde{\bb}, \bgamma) \right)^2 \big| \by, \bX, \tilde{\bb}, \btheta \right] \\
  &= \Var \left[  \tilde{e}_i(\bx_i, \bbeta, \tilde{\bb}, \bgamma) - \frac{1}{n} \sum_{j=1}^{n} \tilde{e}_j(\bx_j, \bbeta, \tilde{\bb}, \bgamma) \big| \by, \bX, \tilde{\bb}, \btheta \right] \\
  &= \Var\left[  \tilde{y}_i - \bar{\tilde{y}} \big| \by, \bX, \tilde{\bb}, \btheta \right],
\end{align*}
we have
\begin{align*}
  \widetilde{\mathrm{RSS}}(\tilde{\bb}, \btheta|\by, \bX)  &= \EE\left[ \sum_{i=1}^{n} \left( \tilde{e}_i(\bx_i, \bbeta, \tilde{\bb}, \bgamma) - \frac{1}{n} \sum_{j=1}^{n} \tilde{e}_j(\bx_j, \bbeta, \tilde{\bb}, \bgamma) \right)^2 | \by, \bX, \tilde{\bb}, \btheta \right] \\
  &= \sum_{i=1}^{n} \Var\left[  \tilde{y}_i - \bar{\tilde{y}} \big| \by, \bX, \tilde{\bb}, \btheta \right].
\end{align*}
However,
\begin{align*}
   \Var\left[  \tilde{y}_i - \bar{\tilde{y}} \big| \by, \bX,\tilde{\bb}, \btheta \right] &= \Var\left[  \tilde{y}_i \big| \by, \bX, \tilde{\bb}, \btheta \right] + \Var\left[ \bar{\tilde{y}} \big| \by, \bX, \tilde{\bb}, \btheta \right] \\
   &- 2 \Cov\left[  \tilde{y}_i, \bar{\tilde{y}} \big| \by, \bX, \tilde{\bb}, \btheta \right] \\
   &= \sigma^{2}(\bx_i, \bbeta, \bphi, \tilde{\bb},\bgamma) + \frac{1}{n^2} \sum_{i=1}^{n} \sigma^{2}(\bx_i, \bbeta, \bphi, \tilde{\bb}, \bgamma) \\
   &- \frac{2}{n} \sigma^{2}(\bx_i, \bbeta, \bphi, \tilde{\bb}, \bgamma) \\
   &= \frac{n-2}{n} \sigma^{2}(\bx_i, \bbeta, \bphi, \tilde{\bb}, \bgamma) + \frac{1}{n^2} \sum_{i=1}^{n} \sigma^{2}(\bx_i, \bbeta, \bphi, \tilde{\bb}, \bgamma).
\end{align*}
Therefore,
\[
  \widetilde{\mathrm{RSS}}(\tilde{\bb}, \btheta|\by, \bX) = (n-1) \frac{1}{n} \sum_{i=1}^{n} \sigma^{2}(\bx_i, \bbeta, \bphi, \tilde{\bb}, \bgamma)
\]
and
\begin{align*}
  R^{2}_{\mathrm{Bayes}} &= \frac{\widetilde{\mathrm{ESS}}(\tilde{\bb}, \btheta| \bX)}{\widetilde{\mathrm{TSS}}(\tilde{\bb}, \btheta|\by, \bX)} \\
  &= \frac{\widetilde{\mathrm{ESS}}(\tilde{\bb}, \btheta| \bX)/(n-1)}{\widetilde{\mathrm{RSS}}(\tilde{\bb}, \btheta|\by, \bX)/(n-1) + \widetilde{\mathrm{ESS}}(\tilde{\bb}, \btheta| \bX)/(n-1)} \\
    & = \frac{ \frac{1}{n-1}\sum_{i=1}^{n} \left( \mu(\bx_i, \bbeta, \tilde{\bb}, \bgamma) - \frac{1}{n} \sum_{j=1}^{n} \mu(\bx_j, \bbeta, \tilde{\bb}, \bgamma) \right)^2 }{ \frac{1}{n} \sum_{i=1}^{n} \sigma^{2}(\bx_i, \bbeta, \bphi, \tilde{\bb}, \bgamma) +  \frac{1}{n-1}\sum_{i=1}^{n} \left( \mu(\bx_i, \bbeta, \tilde{\bb}, \bgamma) - \frac{1}{n} \sum_{j=1}^{n} \mu(\bx_j, \bbeta, \tilde{\bb}, \bgamma) \right)^2 }.
\end{align*}

\cite{gelman2019r} considered the statistic
\[
  S(\by) = \frac{1}{n-1}\sum_{i=1}^{n}\left( y_i - \bar{y} \right)
\]
and argued that, since
\[
 R^{2}_{\mathrm{classic}} = \frac{S(\EE[\by|\bX,\bbeta])}{S(\by)},
\]
a Bayesian alternative for~\eqref{eq:classicalr2} in GLMs with Gaussian or binomial responses is
\[
  R_{\mathrm{Bayes}}^{2} = \frac{\varfit}{\varfit + \varres} = \frac{S\left( \EE[\tilde{\by} | \bX,\btheta] \right)}{ \EE\left[ S(\tilde{\by} - \EE[\tilde{\by} |\bX,\btheta]) \big| \bX,\btheta \right] + S\left( \EE[\tilde{\by} | \bX,\btheta] \right) },
\]
where
\[
  \varfit = S\left( \EE[\tilde{\by} | \bX,\btheta] \right) = \frac{1}{n-1} \sum_{i=1}^{n} \left( \mu(\bx_i, \bbeta) - \frac{1}{n} \sum_{j=1}^{n} \mu(\bx_j, \bbeta) \right)^2
\]
and
\begin{align*}
  \varres &= \EE\left[ S\left(\tilde{\by} - \EE[\tilde{\by} |\bX,\btheta]\right) \big| \bX,\btheta \right] \\
  &= \frac{1}{n-1} \sum_{i=1}^{n} \EE\left[  \left( \left( \tilde{y}_i  - \mu(\bx_i, \bbeta) \right) - \frac{1}{n} \sum_{j=1}^{n} \left(\tilde{y}_j - \mu(\bx_j, \bbeta) \right) \right)^2 \big| \bX,\btheta \right].
\end{align*}

\section{Decomposition of residual sum of squares}
\label{sec:partialdecomp}

For the reduced model, the posterior predictive residuals are given by
\[
  \tilde{e}_{0,i}(\bx_i, \bbeta)= \tilde{y}_i - \mu_0(\bx_i, \bbeta), \quad i=1,\ldots,n,
\]
and its conditional expectation under the full model is
\[
  \EE\left[  \tilde{e}_{0,i}(\bx_i, \bbeta) \big| \by, \bX, \tilde{\bb}, \btheta \right] = \mu(\bx_i, \bbeta, \tilde{\bb}, \bgamma) - \mu_0(\bx_i, \bbeta).
\]
They are related to the posterior predictive residuals of the full model with
\[
  \tilde{e}_{0,i}(\bx_i, \bbeta) = \tilde{e}_i(\bx_i, \bbeta, \tilde{\bb}, \bgamma) + \left( \mu(\bx_i, \bbeta, \tilde{\bb}, \bgamma) - \mu_0(\bx_i, \bbeta) \right)
\]
and hence
\begin{align*}
  \sum_{i=1}^{n} &\left( \tilde{e}_{0,i}(\bx_i, \bbeta) - \frac{1}{n} \sum_{j=1}^{n} \tilde{e}_{0,j}(\bx_j, \bbeta) \right)^2 = \sum_{i=1}^{n} \left( \tilde{e}_{i}(\bx_i, \bbeta, \tilde{\bb}, \bgamma) - \frac{1}{n} \sum_{j=1}^{n} \tilde{e}_{j}(\bx_j, \bbeta, \tilde{\bb}, \bgamma) \right)^2\\
  &+ \sum_{i=1}^{n} \left( \left( \mu(\bx_i, \bbeta, \tilde{\bb}, \bgamma) - \mu_0(\bx_i, \bbeta) \right) - \frac{1}{n} \sum_{j=1}^{n} \left( \mu(\bx_j, \bbeta, \tilde{\bb}, \bgamma) - \mu_0(\bx_j, \bbeta) \right) \right)^2 \\
  &+2  \sum_{i=1}^{n} \left( \tilde{e}_i(\bx_i, \bbeta, \tilde{\bb}, \bgamma) - \frac{1}{n} \sum_{j=1}^{n} \tilde{e}_j(\bx_j, \bbeta, \tilde{\bb}, \bgamma) \right)  \\
   & \hspace*{2cm} \times \left( \left( \mu(\bx_i, \bbeta, \tilde{\bb}, \bgamma) - \mu_0(\bx_i, \bbeta, \bgamma) \right) - \frac{1}{n} \sum_{j=1}^{n} \left( \mu(\bx_j, \bbeta, \tilde{\bb}, \bgamma) - \mu_0(\bx_j, \bbeta) \right) \right).
\end{align*}
Since $\EE\left[  \tilde{y}_i \big| \by, \bX, \tilde{\bb}, \btheta \right] = m(\bx_i,\bbeta, \tilde{\bb})$, we have
\begin{align*}
  \EE \left[ \left( \tilde{e}_i(\bx_i, \bbeta, \tilde{\bb}, \bgamma) - \frac{1}{n} \sum_{j=1}^{n} \tilde{e}_j(\bx_j, \bbeta, \tilde{\bb}, \bgamma) \right)  \big| \by, \bX, \tilde{\bb}, \btheta \right] = 0
\end{align*}
and hence the posterior predictive residual sum of squares for the reduced model is
\begin{align*}
  \widetilde{\mathrm{RSS}}_0(\tilde{\bb}, \btheta| \by, \bX) &= \EE \left[ \sum_{i=1}^{n} \left( \tilde{e}_{0,i}(\bx_i, \bbeta) - \frac{1}{n} \sum_{j=1}^{n} \tilde{e}_{0,j}(\bx_j, \bbeta) \right)^2 \big| \by, \bX, \tilde{\bb}, \btheta \right] \\
  &= \widetilde{\mathrm{RSS}}(\tilde{\bb}, \btheta| \by, \bX) \\
  &+ \sum_{i=1}^{n} \left( \left( \mu(\bx_i, \bbeta, \tilde{\bb}, \bgamma) - \mu_0(\bx_i, \bbeta) \right) - \frac{1}{n} \sum_{j=1}^{n} \left( \mu(\bx_j, \bbeta, \tilde{\bb}, \bgamma) - \mu_0(\bx_j, \bbeta) \right) \right)^2.
\end{align*}


\end{document}